\documentclass[12pt]{article}

\usepackage{amsfonts}
\usepackage{amsmath}
\usepackage{amssymb}
\usepackage{bm}
\usepackage{booktabs}  
\usepackage{caption}
\usepackage{xcolor}
\usepackage{diagbox}
\usepackage{enumerate}
\usepackage{float}
\usepackage{graphics}
\usepackage{lineno}
\usepackage{listings}
\usepackage[bottom=1in,top=1in,left=0.8in,right=0.8in]{geometry}
\usepackage{mathtools}
\usepackage{mdframed}
\usepackage{multicol}
\usepackage{multirow}
\usepackage{natbib}
\usepackage{pdfpages}
\usepackage{rotating}
\usepackage{setspace}
\usepackage{subcaption}
\usepackage{threeparttable}  
\usepackage[normalem]{ulem}
\usepackage{url}
\usepackage{soul}

\usepackage{tikz}
\usetikzlibrary{shapes,decorations,arrows,calc,arrows.meta,fit,positioning}
\tikzset{
	-Latex,auto,node distance=1 cm and 1 cm,semithick,
	state/.style={ellipse,draw,minimum width=0.7 cm},
	point/.style={circle,draw,inner sep=0.04cm,fill,node contents={}},
	bidirected/.style={Latex-Latex,dashed},
	el/.style={inner sep=2pt,align=left,sloped}
}

\newcommand{\longoverbrace}[2]{\overbrace{#1}^{\text{\hbox to 0cm{\hss #2 \hss}}}}
\newcommand{\longunderbrace}[2]{\underbrace{#1}_{\text{\hbox to 0cm{\hss #2 \hss}}}}

\newtheorem{Thm}{Theorem}

\newtheorem{Pro}{Proposition}

\graphicspath{{figure/}}

\title{Causal Discovery with Heterogeneous Observational Data}
\author{Fangting Zhou$^{1,2}$, Kejun He$^{2,\ast}$, Yang Ni$^{1,\ast}$ \\
	$^{1}$Department of Statistics, Texas A\&M University, College Station, Texas, U.S.A. \\
	$^{2}$Institute of Statistics and Big Data, Renmin University of China, Beijing, China \\}
\date{}

\begin{document}
	
\maketitle

\begin{abstract}
We consider the problem of causal discovery (structure learning) from heterogeneous observational data. Most existing methods assume a homogeneous sampling scheme, which leads to misleading conclusions when violated in many applications. To this end, we propose a novel approach that exploits data heterogeneity to infer possibly cyclic causal structures from causally insufficient systems. The core idea is to model the direct causal effects as functions of exogenous covariates that properly explain data heterogeneity. We investigate structure identifiability properties of the proposed model. Structure learning is carried out in a fully Bayesian fashion, which provides natural uncertainty quantification. We demonstrate its utility through extensive simulations and a real-world application.
\end{abstract}

\section{INTRODUCTION} \label{introduction}
Causal discovery is a central task in various fields including social science, artificial intelligence, and systems biology. While randomized controlled trials are the gold standard to establish causality, they can be too costly, unethical, or impossible to carry out. For example, recovering gene regulatory networks through gene knockout would be too expensive to scale whereas observational genomic data are considerably easier to obtain with next-generation sequencing technologies and have become widely available. Many causal discovery methods, therefore, attempt to discover causality from purely observational data.

\paragraph{Related work} One prominent approach in presenting and learning causality is to use a structural equation model (SEM) and the associated causal graph \citep{pearl1998graphs}. The recursive linear Gaussian SEM is among the most popular ones although the associated causal directed acyclic graph is only identifiable up to Markov equivalence classes \citep{verma1990equivalence}. In order to uniquely identify causal structures with observational data, additional distributional assumptions have been made in prior works including the linear non-Gaussian model \citep{shimizu2006linear}, the nonlinear additive noise model \citep{hoyer2008nonlinear}, and the linear Gaussian model with equal error variances \citep{peters2014identifiability}. A common thread of these methods is that they assume that the causal graph is acyclic and there are no unmeasured confounders (also known as causal sufficiency). However, directed cycles and confounders are very common in practice. For example, feedback loops (directed cycles) are common regulatory motifs in biological signaling systems \citep{brandman2008feedback}. As for confounders, gene regulation is known to be affected by many factors such as epigenetic modification \citep{portela2010epigenetic}, which may not be available with gene expression data.

To allow for cycles, non-recursive SEMs have been developed and proven to be identifiable for linear non-Gaussian models \citep{lacerda2008discovering} and the nonlinear additive noise models \citep{mooij2011causal}. In the presence of unmeasured confounders, linear non-Gaussian SEMs have received lots of attention: various models have been proposed and shown to be structurally identifiable under the respective confounding assumptions \citep{hoyer2008estimation, chen2012causal, shimizu2014bayesian, salehkaleybar2020learning}. Nevertheless, none of the aforementioned methods explicitly deal with and provide identifiability guarantees for graphs with both cycles and confounders. Although \cite{hyttinen2012learning} and \cite{forre2018constraint} provided learning algorithms for general SEMs (allowing cycles, confounders, and nonlinearity), full structure can only be recovered with interventional data, which is quite different from the purely observational setting considered in this paper.

Furthermore, all the aforementioned methods assume independent and identically distributed (iid) observations, which may be violated in many applications. For example, cancer is known to be a genetically heterogeneous disease and, therefore, cancer genomic data exhibit great heterogeneity. Methods that ignore such heterogeneity can perform poorly as we will see in our experiments. Recently, \cite{ni2019bayesian, huang2020causal, saeed2020causal} explicitly addressed the heterogeneity issue by incorporating covariates or using a latent mixture model but their models are acyclic and do not account for unmeasured confounders.
For heterogeneous data, \cite{peters2016causal} is able to identify the causal relationships if they are invariant across environment. By contrast, we assume the causal mechanism varies with the environment, which is motivated by biological systems. 
For instance, critical gene pathways have been known to change during the course of cancer development \citep{moustakas2007, huang2009}; hence, in this case, causal variance rather than invariance is a more appropriate assumption.
\cite{mooij2020joint} proposed 
a flexible joint causal inference (JCI) framework, which allows the causal mechanism to vary. However, it did not perform well on our specific real data examples (see Section 4.2).

In this paper, we propose a novel method for \textbf{C}ausal discovery with \textbf{H}eterogeneous \textbf{O}bservational \textbf{D}ata (CHOD). Importantly, we do not restrict our model to be acyclic and do not assume causal sufficiency. By exploiting the data heterogeneity via exogenous covariates, we provide sufficient conditions under which CHOD is structurally identifiable in (i) causally insufficient bivariate cyclic graphs, (ii) causally insufficient multivariate acyclic graphs, and (iii) causally sufficient multivariate cyclic graphs. Our method is among the first model-based causal discovery methods to identify unique causal graphs with both cycles and confounders in purely observational data without prior domain knowledge. Extensive simulation experiments and a real-world application support the identifiability of our method and demonstrate its superiority in handling heterogeneous data through comparison with state-of-the-art alternatives.

\section{PRELIMINARIES OF CAUSAL DISCOVERY} \label{preliminaries}
Let $\bm{X} = (X_1, \dots, X_p)^T$ be a $p$-dimensional random vector. We represent the causal structure as well as the joint observational distribution of $\bm{X}$ by a linear SEM, $\bm{X} = \bm{B} \bm{X} + \bm{\mathcal{E}}$, with direct causal effects $\bm{B} = [b_{j \ell}]\in \mathbb{R}^{p\times p}$ and random noises $\bm{\mathcal{E}} = [\varepsilon_j]\in \mathbb{R}^{p}$. If $b_{j \ell} \neq 0$, then $X_\ell$ is a direct cause of $X_j$. We assume $\bm{\mathcal{E}}$ to be centered Gaussian with covariance $\bm{S} = [\sigma_{j \ell}]$. When there are no unmeasured confounders (i.e., hidden common causes), the noises are independent of each other and hence $\bm{S}$ is diagonal. However, the presence of confounders would correlate the noises, making the off-diagonal elements of $\bm{S}$ non-zero and resulting in a causally insufficient system.

We use a \emph{mixed graph} $\mathcal{G}_M = (V, E_B, E_D)$ to represent the causal relationships embedded in the SEM, where $V = \{1, \ldots, p\}$ is the set of \emph{nodes} representing $\bm{X}$, $E_B$ is the set of \emph{bidirected edges}, and $E_D$ is the set of \emph{directed edges}; see Figure \ref{illustration} for a few examples. There is a bidirected edge $\ell \leftrightarrow j$ if $\sigma_{j \ell} \neq 0$, and a directed edge $\ell \to j$ if $b_{j \ell} \neq 0$. In the former case, $X_j$ and $X_\ell$ are confounded by at least one hidden common cause, while in the latter case, $X_\ell$ is a direct cause of $X_j$. The graph is acyclic if there does not exist a directed path $k_1 \to k_2 \to \ldots \to k_\ell \to k_1$ that returns a node to itself, otherwise it is called cyclic. Our goal is to identify the edge-induced subgraph $\mathcal{G}=(V,E_D)$ with direct causal relationships $E_D$ among the observed variables $\bm{X}$.

\section{METHOD} \label{model}

\subsection{Proposed Model}
Our key idea to discover causality is to take advantage of the data heterogeneity, which we assume can be explained by some exogenous covariates $\bm{Z} \in \mathbb{R}^q$. The exogenous covariates may be observed (e.g., biomarkers in cancer genomic data) or latent. In the latter case, one can impute the latent covariates by various embedding methods such as t-SNE \citep{maaten2008visualizing} and UMAP \citep{mcinnes2018umap}. Alternatively, latent covariates can be learned simultaneously with our model. For ease of exposition, we first focus our discussion on the case where the exogenous covariate is given (either observed or imputed) and univariate (i.e., $q = 1$), and later briefly discuss the extension to multivariate latent covariates. Specifically, given 
$Z$, we model $\bm{X}$ as a conditionally linear Gaussian SEM,
\begin{align} \label{sem}
\bm{X} = \bm{B}(Z) \bm{X} + \bm{\mathcal{E}}, ~~ \bm{\mathcal{E}} \sim N(\bm{0}, \bm{S}),
\end{align}
where $\bm{B}(Z) = [b_{j \ell}(Z)]: \mathbb{R}\mapsto\mathbb{R}^{p\times p}$ is a matrix-valued function of $Z$, which characterizes the changes of the direct causal effects with respect to $Z$. Because each observation potentially has a different value of covariate $Z$, the direct causal effects $\bm{B}(Z)$ are heterogeneous and observation-specific. Note that for simplicity, we have assumed linearity and Gaussianity in the current formulation. However, extension to nonlinear models is possible with proper basis expansion (e.g., splines) and incorporation of exogenous covariates into the basis coefficients. Moreover, as demonstrated in the experiments and discussion, the Gaussian error is not a crucial assumption of our model. Model \eqref{sem} implies the conditional distribution of $\bm{X}$ given $Z$, 
$$\mathbb{P}(\bm{X}|Z,\bm{B},\bm{S})=\mathrm{det}(\bm{I} - \bm{B}(Z))N((\bm{I}-\bm{B}(Z))\bm{X}|\bm{0}, \bm{S}).$$
When $\bm{B}(Z)$ is constant in $Z$, model \eqref{sem} is reduced to an ordinary linear Gaussian SEM and hence its underlying causal graph $\mathcal{G}$ is not identifiable. However, as we will show later, the causal graph of model \eqref{sem} is in general identifiable.  While most existing methods assume the environment or the exogenous covariate to be discrete and finite (i.e., multiple contexts, domains, or experimental conditions), under our framework the exogenous covariate $Z$ can be either continuous or discrete. To match the case of our real-data application (See Section 4.2) and emphasize the advantage of CHOD, we present $Z$ as a continuous covariate in this paper. Therefore, our method will be particularly useful when the data are heterogeneous but there are no clear predefined discrete groups. Our model is also reminiscent of causal models with soft interventions by viewing $Z$ as intervention that modifies causal effects; however, the key difference is that our model does not assume the knowledge of the interventional targets (i.e., $Z$ could affect all causal effects). In summary, our model explicitly accounts for the heterogeneity of data generating mechanism via the observation-specific direct causal effects $\bm{B}(Z)$, which vary smoothly with covariate $Z$.

\subsection{Causal Structure Identifiability}
Formally, for model-based causal discovery methods, the non-identifiability issue can be seen from the distribution equivalence point of view. Two CHOD models parameterized by $(\bm{B}, \bm{S})$ and $(\bm{B}', \bm{S}')$ are said to be \emph{distribution equivalent} if for any values of $(\bm{B}, \bm{S})$ there exist values of $(\bm{B}', \bm{S}')$ such that $\mathbb{P}(\bm{X} | Z, \bm{B}, \bm{S}) = \mathbb{P}(\bm{X} | Z, \bm{B}', \bm{S}')$ for all $\bm{X}$. Clearly, distribution equivalent models cannot be distinguished from each other by examining their observational distributions. The causal structure is said to be \emph{identifiable} if there do not exist two distribution equivalent causal models such that $\mathcal{G} \neq \mathcal{G}'$.

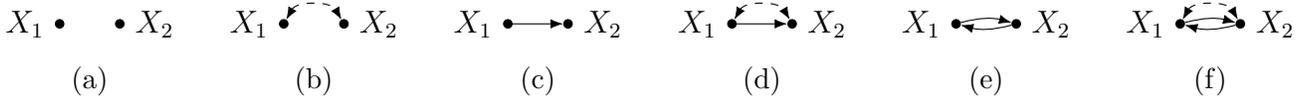
\begin{figure*}[h]
\centering
\begin{subfigure}{0.15\textwidth}
\centering
\begin{tikzpicture}
\node (x) at (0,0) [label=left:$X_1$,point];
\node (y) at (0.8,0) [label=right:$X_2$,point];
\end{tikzpicture}
\caption{}	
\end{subfigure}
\hfill
\begin{subfigure}{0.15\textwidth}
\centering
\begin{tikzpicture}
\node (x) at (0,0) [label=left:$X_1$,point];
\node (y) at (0.8,0) [label=right:$X_2$,point];
\path[bidirected] (x) edge[bend left=60] (y);
\end{tikzpicture}
\caption{}
\end{subfigure}
\hfill
\begin{subfigure}{0.15\textwidth}
\centering
\begin{tikzpicture}
\node (x) at (0,0) [label=left:$X_1$,point];
\node (y) at (0.8,0) [label=right:$X_2$,point];
\path (x) edge (y);
\end{tikzpicture}
\caption{}
\end{subfigure}
\hfill
\begin{subfigure}{0.15\textwidth}
\centering
\begin{tikzpicture}
\node (x) at (0,0) [label=left:$X_1$,point];
\node (y) at (0.8,0) [label=right:$X_2$,point];
\path (x) edge (y);
\path[bidirected] (x) edge[bend left=60] (y);
\end{tikzpicture}
\caption{}
\end{subfigure}
\hfill
\begin{subfigure}{0.15\textwidth}
\centering
\begin{tikzpicture}
\node (x) at (0,0) [label=left:$X_1$,point];
\node (y) at (0.8,0) [label=right:$X_2$,point];
\path (x) edge [bend left=15](y);
\path (y) edge [bend left=15](x);
\end{tikzpicture}
\caption{}
\end{subfigure}
\hfill
\begin{subfigure}{0.15\textwidth}
\centering
\begin{tikzpicture}
\node (x) at (0,0) [label=left:$X_1$,point];
\node (y) at (0.8,0) [label=right:$X_2$,point];
\path (x) edge [bend left=15](y);
\path (y) edge [bend left=15](x);
\path[bidirected] (x) edge[bend left=60] (y);
\end{tikzpicture}
\caption{}
\end{subfigure}
\caption{Mixed graphs. Solid arrows are causal effects and dashed bidirected arrows are confounding effects.}
\label{illustration}
\end{figure*}

Throughout, we make the causal Markov assumption \citep{richardson1996discovery}, i.e., the probability distribution $\mathbb{P}$ respects the Markov property of the causal graph $\mathcal{G}$. Before stating our main results, we first provide an intuition on how the proposed CHOD is identifiable using a toy example. Consider the bivariate graphs shown in Figure \ref{illustration}. We can distinguish graphs (a)--(b) from graphs (c)--(f) because the marginal variance of $X_2$ is independent of $Z$ in graphs (a)--(b) but depends on $Z$ in graphs (c)--(f) through the causal effect $b_{21}(Z)$: $X_1\rightarrow X_2$. Likewise, we can separate graphs (c)--(d) from graphs (e)--(f) by examining the marginal variance of $X_1$ which depends on $Z$ through $X_2\rightarrow X_1$. We may not distinguish (c) from (d) or (e) from (f), but the direct causal relationship between $X_1$ and $X_2$ is determined in either case.

\begin{figure}[h]
\centering
\includegraphics[width=0.45\textwidth]{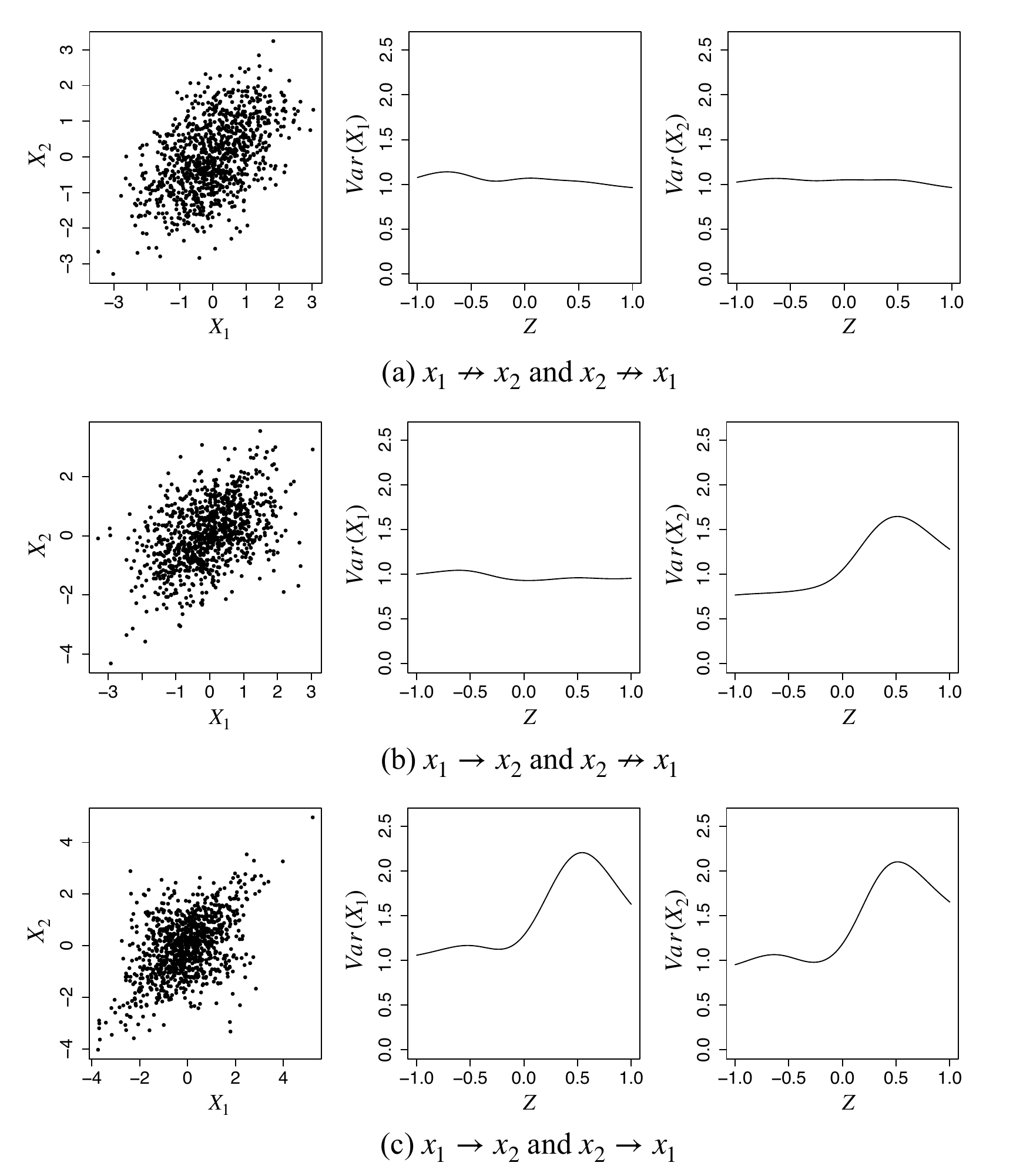}
\caption{Illustration of identifiability with a bivariate toy example.}
\label{bv}
\end{figure}

We further illustrate the identifiability with simulated data from graphs (b), (d), and (f) in Figure \ref{illustration}. Specifically, the exogenous covariates were generated uniformly. Under each graph, the non-zero elements of $\bm{B}(Z)$ were assumed to be $0.5 \sin(\pi Z)$. We set the noise variances to 1 and the correlation coefficients to 0.5 to have confounding effects. The $n = 1000$ data points as well as the marginal variances estimated by kernel methods of the two nodes as functions of $Z$ are depicted in Figure \ref{bv} for these 3 cases, from which the causal relationships between $X_1$ and $X_2$ are intuitively identifiable in the presence of both confounders and cycles: in Figure \ref{bv}(a), both Var$(X_1)$ and Var$(X_2)$ are constant in $Z$ indicating no direct causal link; in Figure \ref{bv}(b), Var$(X_1)$ is constant but Var$(X_2)$ is not constant in $Z$ indicating a direct causal link $X_1\rightarrow X_2$; and in Figure \ref{bv}(c), neither Var$(X_1)$ nor Var$(X_2)$ is constant in $Z$ indicating a cyclic causal link $X_1\rightleftarrows X_2$. 

Now we present the identifiability theories. We first show that in the bivariate case the CHOD models are not distribution equivalent and therefore their causal graphs are identifiable.

\begin{Thm}[Causally Insufficient Bivariate Cyclic Graphs] \label{thm1}
Consider bivariate CHOD models with direct causal effects $[b_{12}(Z), b_{21}(Z)]$ and $[b'_{12}(Z), b'_{21}(Z)]$, respectively. Assume $b_{j \ell}(Z)$ and $b'_{j \ell}(Z)$ are either zero or non-constant functions for all $j \neq \ell \in \{1, 2\}$. Then if the two CHOD models are distribution equivalent, we must have $\mathcal{G} = \mathcal{G}'$.
\end{Thm}

All proofs are provided in Section S1 of the Supplementary Materials. The assumption that $b_{j \ell}(Z)$ is either zero or a non-constant function is not surprising because if non-zero $b_{j \ell}(Z)$ is constant in $Z$, then the proposed model is reduced to an ordinary linear Gaussian SEM, which is known to be non-identifiable. Loosely speaking, Theorem \ref{thm1} states that CHOD is identifiable if $Z$ can help explain the heterogeneity of the data generating mechanism.

Next, we provide sufficient conditions for causally insufficient multivariate acyclic systems and causally sufficient multivariate cyclic systems to be identifiable, and leave the theoretical investigation of causally insufficient multivariate cyclic systems as future work. Denote $pa(j) = \{\ell: \ell \to j \in E_D\}$ as the set of direct causes and $ds(j) = \{\ell: \ell \leftrightarrow \cdots \leftrightarrow j\}$ as the set of nodes connected to $j$ through bidirected arrows.

\begin{Thm}[Causally Insufficient Multivariate Acyclic Graphs] \label{thm2}
Consider the CHOD model in \eqref{sem} restricted to acyclic causal graphs. Assume without loss of generality $(1,\dots,p)$ is a true causal ordering (i.e., $\ell\not\rightarrow j$ if $\ell>j$). If for any node $j$, and any set $S = \{1,\dots,m\}$ such that $pa(j) \not\subset S$, we have $\mathrm{Var}(X_j | \bm{X}_S)$ is a non-constant function of the covariate $Z$, then the causal ordering is identifiable. Moreover, if $pa(j) \cap ds(j) = \emptyset,\forall j$, then the causal graph is identifiable.
\end{Thm}

The proof of Theorem \ref{thm2} first identifies an ordering by recursively finding root variables in acyclic graphs and then identifies the graph structure given the ordering. The assumption on Var$(X_j|\bm{X}_S)$ states that the heterogeneous causal effects do not accidentally become constant in any paths, which is similar in spirit to the causal faithfulness assumption. See Section S1.2 of the Supplementary Materials for more discussions.

We have presented our theorems in their strongest forms, i.e., full structure identifiability. If some of the causal effects do not vary with $Z$, then their identification is not always guaranteed (they may still be identifiable in some graphs via v-structure and Meek rules). This is similar to other causal models. For example, in additive noise models, all causal effects have to be nonlinear for full identification. Those linear causal effects have to rely on v-structure and Meek rules to achieve identification as in our method. In the linear non-Gaussian acyclic model, all but one noises have to be non-Gaussian. Like our model, violation of these assumptions would lead to partial structure identification. We would like to point out though, under our proposed Bayesian learning framework discussed in Section \ref{sec:inf}, we can assess the credibility of inferred edges via posterior inference: edges that have nearly constant causal effects (e.g., if 95\% credible bands of $b_{j\ell}(Z)$, which can be computed from Monte Carlo samples, cover constant functions) are deemed less reliable.

Unlike bivariate graphs, the identifiability results of multivariate cyclic graphs for purely observational data are very sparse in the literature with few exception \citep{lacerda2008discovering}, which assumes causal sufficiency and disjoint cycles. In the following theorem, we also assume the same conditions.

\begin{Thm}[Causally Sufficient Multivariate Cyclic Graphs] \label{thm3}
Consider the CHOD model \eqref{sem}. Assume there are no unmeasured confounders and all cycles are  disjoint. 
The causal graph is generally identifiable\footnote{That is, it is identifiable unless a peculiar condition holds. Because of space, we discuss that condition ($\star$) in the Section S1.3 of the Supplementary Materials.}.
\end{Thm}

Theorems \ref{thm1}--\ref{thm3} assume $Z$ to be univariate and known (observed or imputed). When $\bm{Z}$ is multivariate and unknown, it can be inferred jointly with the causal graph. We provide its identifiability result below and briefly discuss its implementation in the Section S2.1 of the Supplementary Materials. 

\begin{Pro}[Multivariate Latent Exogenous Covariates] \label{prop1}
Assume the vector $\bm{m}(\bm{Z})$ that stacks the non-zero elements of $\bm{B}(\bm{Z})$ is continuous and injective, and $(\bm{m}, \bm{S}) \mapsto \mathbb{P}(\bm{X}|\bm{m},\bm{S})$ is continuous and injective in $\bm{m}$ given $\mathcal{G}$. Then the latent exogenous covariates are identifiable up to a monotone transformation.
\end{Pro}

Proposition \ref{prop1} shows that the \emph{relative} position of the latent covariates can be identified, which is useful in sorting observations (see many prominent examples of trajectory inference in single-cell genomic studies \citep{saelens2019comparison}). It can be also viewed as an embedding and dimension reduction tool wrapped in a causal model because the dimension of $\bm{Z}$ is typically much smaller than $\bm{X}$. The condition of Proposition \ref{prop1} that assumes $\bm{m}(\bm{Z})$ to be injective as vector-valued functions should not be interpreted as a requirement that each individual function has to be injective. For example, if $b_{12}(Z) = (Z+1)^2$ and $b_{21}(Z) = Z^2$, neither is injective but the resulting $\bm{m}(Z) = [b_{12}(Z), b_{21}(Z)]$ is injective. In our formulation where the causal structure is identifiable, the injectivity assumption on the density function is equivalent to the identifiability of causal effects or model parameters given the causal structure. The causal effect identifiability itself is an interesting but challenging task. For linear Gaussian SEMs, it is well-known that causal effects are identifiable without confounders. With confounders, \cite{drton2011global} showed that the acyclic mixed graph needs to be a simple graph. As is evident from the proofs of Theorems \ref{thm2}\&\ref{thm3}, the causal effects under the corresponding assumptions are indeed identifiable. In a related work, \cite{salehkaleybar2020learning} showed that the causal effects in the presence of latent confounders are identifiable with mild structure assumptions in the non-Gaussian setting. This paper focuses on investigating causal structure identifiability; establishing causal effect identifiability theory for the causally insufficient multivariate cyclic graphs will be an interesting future work.

\subsection{Bayesian Structure Learning} \label{sec:inf}

We learn the causal structure through a Bayesian approach by assigning priors on the space of graphs and model parameters. We model the direct causal effects by cubic B-splines with equally spaced knots, $b_{j \ell}(Z) = \sum_{k = 1}^K \beta_{j \ell k} \phi_k(Z)$ where $\{\phi_k(Z)\}_{k=1}^K$ are $K$ spline bases. To encourage graph sparsity, a spike-and-slab prior is assigned to the vector $\bm{\beta}_{j \ell} = (\beta_{j \ell 1}, \ldots, \beta_{j \ell K})^T$,
\begin{align*}
\mathbb{P}(\bm{\beta}_{j \ell}|r_{j \ell}, \tau) = (1 - r_{j \ell}) \delta_{\bm{0}}(\bm{\beta}_{j \ell}) + r_{j \ell} N(\bm{\beta}_{j \ell}|\bm{0}, \tau \bm{I}),
\end{align*}
where $\delta_{\bm{0}}(\cdot)$ is a point mass at vector zero and $r_{j \ell}$ is a binary edge indicator. By construction, $r_{j \ell} = 0$ if and only if $\bm{\beta}_{j \ell} = \bm{0}$ (equivalently, $\ell \not\rightarrow j$ and $b_{j\ell}(Z) \equiv 0$). We assume independent beta-Bernoulli priors with $r_{j \ell}\sim\mathbb{P}(r_{j \ell}|\pi) = \mathrm{Bernoulli}(r_{j \ell}|\pi)$ and $\pi \sim \mathbb{P}(\pi)=\mathrm{beta}(\pi|a, b)$. We place conjugate inverse-gamma prior on $\tau\sim \mathbb{P}(\tau)= IG(\tau|\alpha, \beta)$ and inverse-Wishart prior on the covariance matrix $\bm{S} \sim \mathbb{P}(\bm{S})=IW(\bm{S}|\bm{\Psi}, v)$. If a sparse estimation of confounding effects is desired, selection or shrinkage priors can be assigned to $\bm{S}$, which we do not pursue in this paper.

Let $\mathcal{D} = \{(\bm{x}_i, z_i), i \in 1,\dots,n \}$ be $n$ realizations of $(\bm{X}, Z)$. Denote $\bm{\beta} = [\beta_{j\ell k}]$ and $\bm{r} = [r_{j\ell}]$. The joint posterior distribution is then given by
\begin{align*}
\mathbb{P}(\bm{\beta}, \bm{S}, \bm{r}, \pi, \tau | \mathcal{D}) & \propto \mathbb{P}(\bm{\beta} | \bm{r}, \tau) \mathbb{P}(\bm{S}) \mathbb{P}(\bm{r} | \pi) \mathbb{P}(\pi) \mathbb{P}(\tau) \\ 
& \qquad \times \textstyle\prod_{i = 1}^n \mathbb{P}(\bm{x}_i | z_i, \bm{B}(z_i), \bm{S}),
\end{align*}
where $\mathbb{P}(\bm{\beta} | \bm{r}, \tau) = \prod_{j, \ell} \mathbb{P}(\bm{\beta}_{j \ell} | r_{j \ell}, \tau)$ and $\mathbb{P}(\bm{r} | \pi) = \prod_{j, \ell} \mathbb{P}(r_{j \ell} | \pi)$.
The posterior distribution is not analytically available; we use Markov chain Monte Carlo (MCMC) to approximate it with the sampling steps detailed in Section S2 of Supplementary Materials.

The per-iteration computational complexity of sampling is $O(np^5)$\footnote{Sampling $\bm{r}$ and $\bm{\beta}$ requires $O(p^2)$ numbers of likelihood evaluation and each likelihood evaluation is $O(np^3)$.}. This is a general learning algorithm that includes possible cycles and confounders. It can be simplified if there are no cycles and/or confounders. For example, for acyclic graphs, the spline coefficients $\bm{\beta}$ can be integrated out to improve MCMC mixing. Upon the completion of MCMC, the causal structure can be summarized by thresholding the estimated posterior probability of inclusion $\mathbb{P}(r_{j \ell} = 1 | \mathcal{D}) \approx 1 / M \sum_{m = 1}^M I(r_{j \ell}^{(m)} = 1)$ at 0.5, where the superscript $(m)$ indexes the Monte Carlo samples.

\section{EXPERIMENTS} \label{experiments}
We use extensive simulations as well as a real-world application with known cyclic causal graphs to evaluate the proposed method, CHOD. Throughout, we set the hyperparameters as non-informative ones with $\bm{\Psi} = \bm{I}$, $v = p$, $a = b = 0.5$, $\alpha = \beta = 0.01$, and $K = 10$, which performed well in all experiments considered. We ran MCMC for 2000 iterations, discarded the first 1000 iterations as burn-in, and retained every 5th iteration after burn-in as posterior samples. We evaluated the graph structure recovery accuracy by calculating true positive rate (TPR), false discovery rate (FDR), and Matthew's correlation coefficient (MCC) based on 50 repetitions in simulations. MCC ranges over $[-1, 1]$ with 1 indicating perfect graph recovery.

\subsection{Simulations}
Since most existing causal discovery methods assume acyclic graphs and/or causal sufficiency with some exceptions that allow either cycles or confounders but usually not both, in order to maximize the fairness of comparison, we conducted simulations under three scenarios: when the simulation truths are cyclic graphs with confounders, acyclic graphs with confounders, and cyclic graphs without confounders, respectively. The first scenario is the most general one, which has been the focus of this paper, while the second and third scenarios are designed for fairness and hence are briefly discussed in the main text with details provided in the Section S3.1 of the Supplementary Materials. Note that our general algorithm accommodates all those three settings. We first considered data generated from our proposed model and then considered various model misspecifications in terms of non-Gaussian errors, different confounding effects, varying degrees of heterogeneity, and unobserved covariates. 

\paragraph{Data generating mechanism} We considered sample size $n \in \{125, 250, 500, 1000\}$ and the number of nodes $p \in \{10, 25, 50\}$. Exogenous covariates were simulated from the uniform distribution $U(-1, 1)$. True causal graphs were generated as Erd\H{o}s-R\'{e}nyi random graph with edge probability $1 / p$ (plotted in Section S3 of the Supplementary Materials). When assumed acyclic in the second scenario, the graph is constrained to have no directed cycles. Given the true structure, non-zero direct causal effects were randomly chosen from $f(Z) = 0.8 Z$, $g(Z) = 0.9 \cos(\pi Z)$, or $h(Z) = 0.9 \tanh(\pi Z)$. We set the diagonal elements of $\bm{S}$ to 1. We generated the off-diagonal entries of $\bm{S}$ randomly from $U(-1, 1)$ in scenarios where there are unmeasured confounders, subject to $\bm{S}$ being positive-definite. Observations were then generated from model \eqref{sem}.

\paragraph{Scenario 1: cyclic graphs with confounders} 
To the best of our knowledge, methods that can deal with both cycles and confounders in purely observational data are uncommon. We compared CHOD with two state-of-the-art acyclic causal discovery methods with confounders: RFCI \citep{colombo2012learning} and RICA \citep{salehkaleybar2020learning}, and two state-of-the-art cyclic causal discovery methods without confounders: LiNG \citep{lacerda2008discovering} and ANM \citep{mooij2011causal}. RICA and LiNG are based on linear non-Gaussian models, while ANM uses nonlinear additive noise models. RFCI imposes no distributional assumptions and outputs a graph containing both directed and bidirected edges (or edges with indeterminate directions). The results are summarized in Table \ref{latent}. As expected, CHOD was the only approach that could recover the true graph well under this general heterogeneous simulation setting where both cycles and confounders are present. For example, the MCC for all the competing methods was uniformly low for all $(n,p)$ whereas the MCC of the proposed CHOD was always substantially higher and improved as sample size increased as expected.

\begin{table*}[h]
\caption{Simulation Scenario 1. Average operating characteristics over 50 repetitions. The standard deviation for each statistic is given within parentheses. The best performance is shown in boldface.}
\resizebox{\textwidth}{!}{
\centering
\begin{tabular}{cccccccccc}
\toprule
\multirow{2}{*}{$n = 125$} & \multicolumn{3}{c}{$p = 10$} & \multicolumn{3}{c}{$p = 25$} & \multicolumn{3}{c}{$p = 50$} \cr
\cmidrule(lr){2-4} \cmidrule(lr){5-7} \cmidrule(lr){8-10}
& TPR & FDR & MCC & TPR & FDR & MCC & TPR & FDR & MCC \cr
\midrule
CHOD & 0.662 (0.065) & \textbf{0.253 (0.092)} & \textbf{0.644 (0.077)} & 0.662 (0.062) & \textbf{0.385 (0.080)} & \textbf{0.590 (0.094)} & 0.608 (0.082) & \textbf{0.385 (0.077)} & \textbf{0.590 (0.061)} \cr
LiNG & \textbf{0.913 (0.088)} & 0.863 (0.012) & 0.104 (0.073) & \textbf{0.860 (0.057)} & 0.953 (0.003) & 0.023 (0.034) & \textbf{0.802 (0.059)} & 0.979 (0.002) & 0.007 (0.021) \cr
ANM & 0.093 (0.069) & 0.879 (0.091) & 0.004 (0.083) & 0.002 (0.006) & 0.988 (0.035) & 0.009 (0.015) & 0.002 (0.006) & 0.972 (0.043) & 0.004 (0.022) \cr		
RFCI & 0.174 (0.069) & 0.742 (0.082) & 0.113 (0.071) & 0.227 (0.035) & 0.715 (0.046) & 0.200 (0.041) & 0.046 (0.025) & 0.943 (0.026) & 0.030 (0.025) \cr	
RICA & 0.470 (0.130) & 0.895 (0.029) & 0.051 (0.009) & 0.566 (0.104) & 0.947 (0.009) & 0.037 (0.045) & 0.485 (0.044) & 0.978 (0.002) & 0.006 (0.013) \cr
\midrule
\multirow{2}{*}{$n = 250$} & \multicolumn{3}{c}{$p = 10$} & \multicolumn{3}{c}{$p = 25$} & \multicolumn{3}{c}{$p = 50$} \cr
\cmidrule(lr){2-4} \cmidrule(lr){5-7} \cmidrule(lr){8-10}
& TPR & FDR & MCC & TPR & FDR & MCC & TPR & FDR & MCC \cr
\midrule
CHOD & 0.804 (0.081) & \textbf{0.162 (0.088)} & \textbf{0.768 (0.075)} & 0.698 (0.065) & \textbf{0.286 (0.090)} & \textbf{0.662 (0.056)} & 0.680 (0.083) & \textbf{0.340 (0.081)} & \textbf{0.644 (0.086)} \cr
LiNG & \textbf{0.880 (0.084)} & 0.870 (0.010) & 0.064 (0.068) & \textbf{0.834 (0.099)} & 0.955 (0.005) & 0.007 (0.051) & \textbf{0.798 (0.047)} & 0.979 (0.001) & 0.004 (0.017) \cr
ANM & 0.077 (0.056) & 0.910 (0.057) & 0.024 (0.058) & 0.009 (0.013) & 0.937 (0.048) & 0.010 (0.033) & 0.003 (0.029) & 0.964 (0.012) & 0.004 (0.002) \cr		
RFCI & 0.201 (0.050) & 0.739 (0.045) & 0.123 (0.047) & 0.280 (0.033) & 0.744 (0.027) & 0.203 (0.030) & 0.009 (0.030) & 0.907 (0.027) & 0.069 (0.028) \cr	
RICA & 0.463 (0.127) & 0.890 (0.033) & 0.033 (0.099) & 0.481 (0.120) & 0.955 (0.011) & 0.001 (0.051) & 0.472 (0.065) & 0.978 (0.003) & 0.008 (0.019) \cr
\midrule		
\multirow{2}{*}{$n = 500$} & \multicolumn{3}{c}{$p = 10$} & \multicolumn{3}{c}{$p = 25$} & \multicolumn{3}{c}{$p = 50$} \cr
\cmidrule(lr){2-4} \cmidrule(lr){5-7} \cmidrule(lr){8-10}
& TPR & FDR & MCC & TPR & FDR & MCC & TPR & FDR & MCC \cr
\midrule
CHOD & 0.849 (0.073) & \textbf{0.152 (0.073)} & \textbf{0.813 (0.073)} & 0.813 (0.063) & \textbf{0.268 (0.090)} & \textbf{0.768 (0.094)} & \textbf{0.804 (0.075)} & \textbf{0.259 (0.057)} & \textbf{0.768 (0.061)} \cr
LiNG & \textbf{0.917 (0.099)} & 0.867 (0.013) & 0.093 (0.083) & \textbf{0.875 (0.073)} & 0.952 (0.004) & 0.034 (0.039) & 0.788 (0.063) & 0.979 (0.002) & 0.002 (0.022) \cr
ANM & 0.100 (0.066) & 0.898 (0.071) & 0.026 (0.074) & 0.009 (0.013) & 0.917 (0.058) & 0.009 (0.041) & 0.004 (0.008) & 0.961 (0.015) & 0.003 (0.018) \cr		
RFCI & 0.274 (0.045) & 0.734 (0.050) & 0.147 (0.052) & 0.320 (0.043) & 0.772 (0.030) & 0.197 (0.039) & 0.124 (0.027) & 0.901 (0.018) & 0.086 (0.022) \cr	
RICA & 0.413 (0.126) & 0.903 (0.033) & 0.078 (0.104) & 0.491 (0.094) & 0.954 (0.009) & 0.006 (0.040) & 0.489 (0.082) & 0.977 (0.004) & 0.015 (0.025) \cr
\midrule		
\multirow{2}{*}{$n = 1000$} & \multicolumn{3}{c}{$p = 10$} & \multicolumn{3}{c}{$p = 25$} & \multicolumn{3}{c}{$p = 50$} \cr
\cmidrule(lr){2-4} \cmidrule(lr){5-7} \cmidrule(lr){8-10}
& TPR & FDR & MCC & TPR & FDR & MCC & TPR & FDR & MCC \cr
\midrule
CHOD & \textbf{0.912 (0.081)} & \textbf{0.142 (0.069)} & \textbf{0.840 (0.083)} & \textbf{0.894 (0.072)} & \textbf{0.245 (0.064)} & \textbf{0.813 (0.062)} & \textbf{0.849 (0.093)} & \textbf{0.251 (0.062)} & \textbf{0.786 (0.079)} \cr
LiNG & 0.897 (0.108) & 0.866 (0.014) & 0.088 (0.089) & 0.836 (0.058) & 0.954 (0.003) & 0.013 (0.030) & 0.814 (0.058) & 0.978 (0.001) & 0.011 (0.020) \cr
ANM & 0.127 (0.059) & 0.908 (0.042) & 0.037 (0.055) & 0.010 (0.006) & 0.903 (0.006) & 0.024 (0.006) & 0.003  (0.016) & 0.962 (0.017) & 0.009 (0.017) \cr		
RFCI & 0.303 (0.043) & 0.754 (0.034) & 0.139 (0.042) & 0.350 (0.028) & 0.792 (0.017) & 0.191 (0.022) & 0.159 (0.022) & 0.897 (0.013) & 0.101 (0.017) \cr	
RICA & 0.447 (0.142) & 0.893 (0.037) & 0.044 (0.114) & 0.506 (0.073) & 0.951 (0.007) & 0.018 (0.031) & 0.483 (0.078) & 0.977 (0.004) & 0.014 (0.023) \cr
\bottomrule					
\end{tabular}}
\label{latent}
\end{table*}

\paragraph{Scenario 2: acyclic graphs with confounders} In addition to RICA and RFCI, in this scenario, we also compared CHOD with CAM \citep{buhlmann2014cam},
GDS \citep{peters2014causal}, and RESIT \citep{peters2014causal} as benchmarks although they are not designed for causal discovery in the presence of confounders. These three methods are based on nonlinear additive noise model. Moreover, we combined several bivariate causal discovery methods with CAM as suggested in \citet{tagasovska2020distinguishing} by first using CAM to learn a Markov equivalence class and then using IGCI \citep{janzing2010causal}, EMD \citep{chen2014causal}, and bQCD \citep{tagasovska2020distinguishing} to orient edges. These three bivariate causal discovery methods are based on asymmetry between the cause and the effect in terms of certain complexity metrics. Results are summarized in Section S3.1 of the Supplementary Materials. In summary, CHOD outperformed all the competing methods: the MCC of CHOD ranged from 0.6 to 0.9 whereas the competing methods had MCC typically $<$0.4. Furthermore, in Section S3.1 of the Supplementary Materials, we considered more comparisons with methods that incorporated the covariate $Z$ as an additional node in the causal graph (similar in spirit to the JCI framework \citep{mooij2020joint}). However, these additional comparisons did not show significantly better graph recovery. 

\paragraph{Scenario 3: cyclic graphs without confounders} We compared CHOD with LiNG and ANM. The results are summarized in Section S3.1 of the Supplementary Materials. As in the first two scenarios, CHOD performed significantly better by exploiting the data heterogeneity.

\begin{figure*}[h]
\centering
\begin{subfigure}[b]{0.3 \textwidth}
\centering
\includegraphics[width = 0.9 \textwidth]{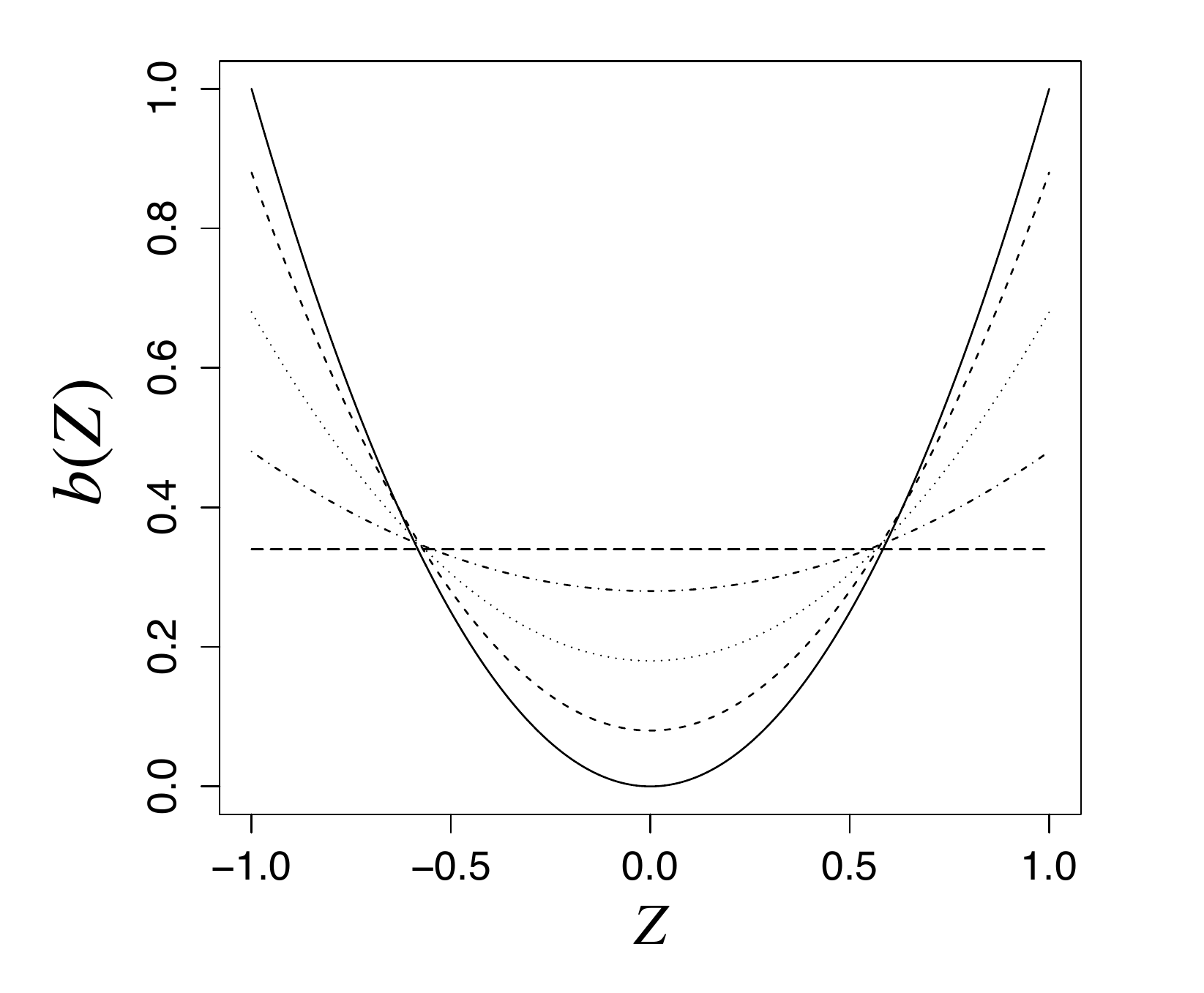}
\caption{Direct causal effect functions.}
\end{subfigure}
\begin{subfigure}[b]{0.3 \textwidth}
\centering
\includegraphics[width = 0.8 \textwidth]{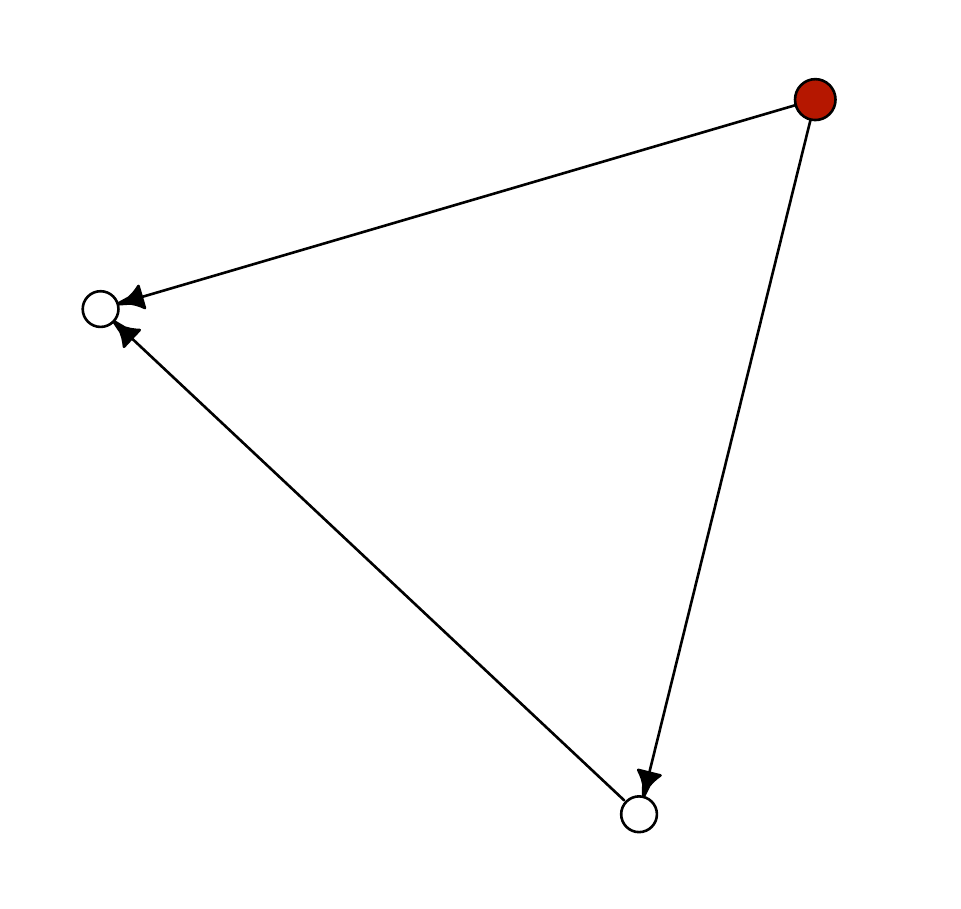}
\caption{The three-node graph.}
\end{subfigure}
\begin{subfigure}[b]{0.3 \textwidth}
\centering
\includegraphics[width = 0.8 \textwidth]{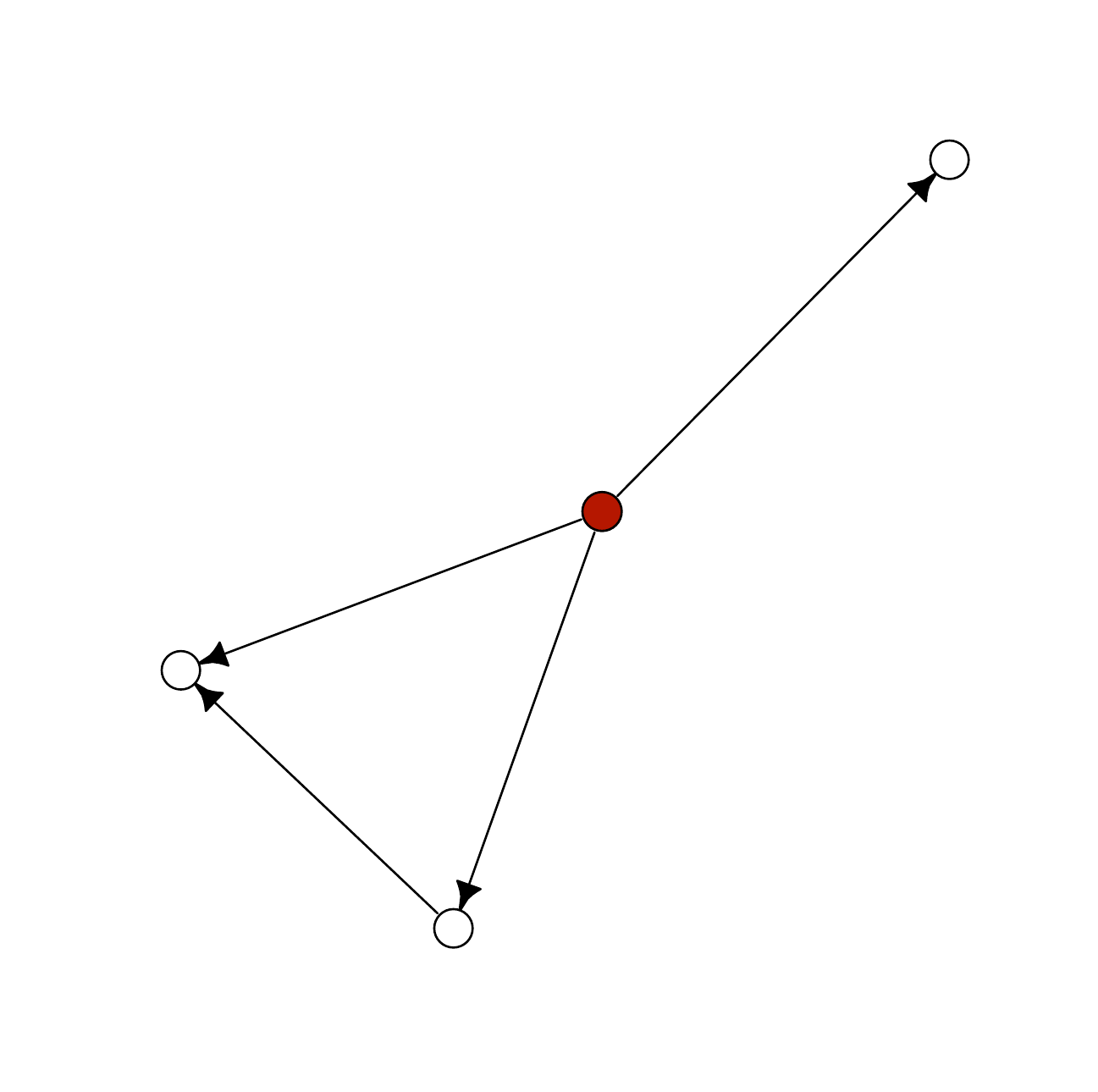}
\caption{The four-node graph.}
\end{subfigure}
	
\begin{subfigure}[b]{0.46 \textwidth}
\centering
\includegraphics[width = 1 \textwidth]{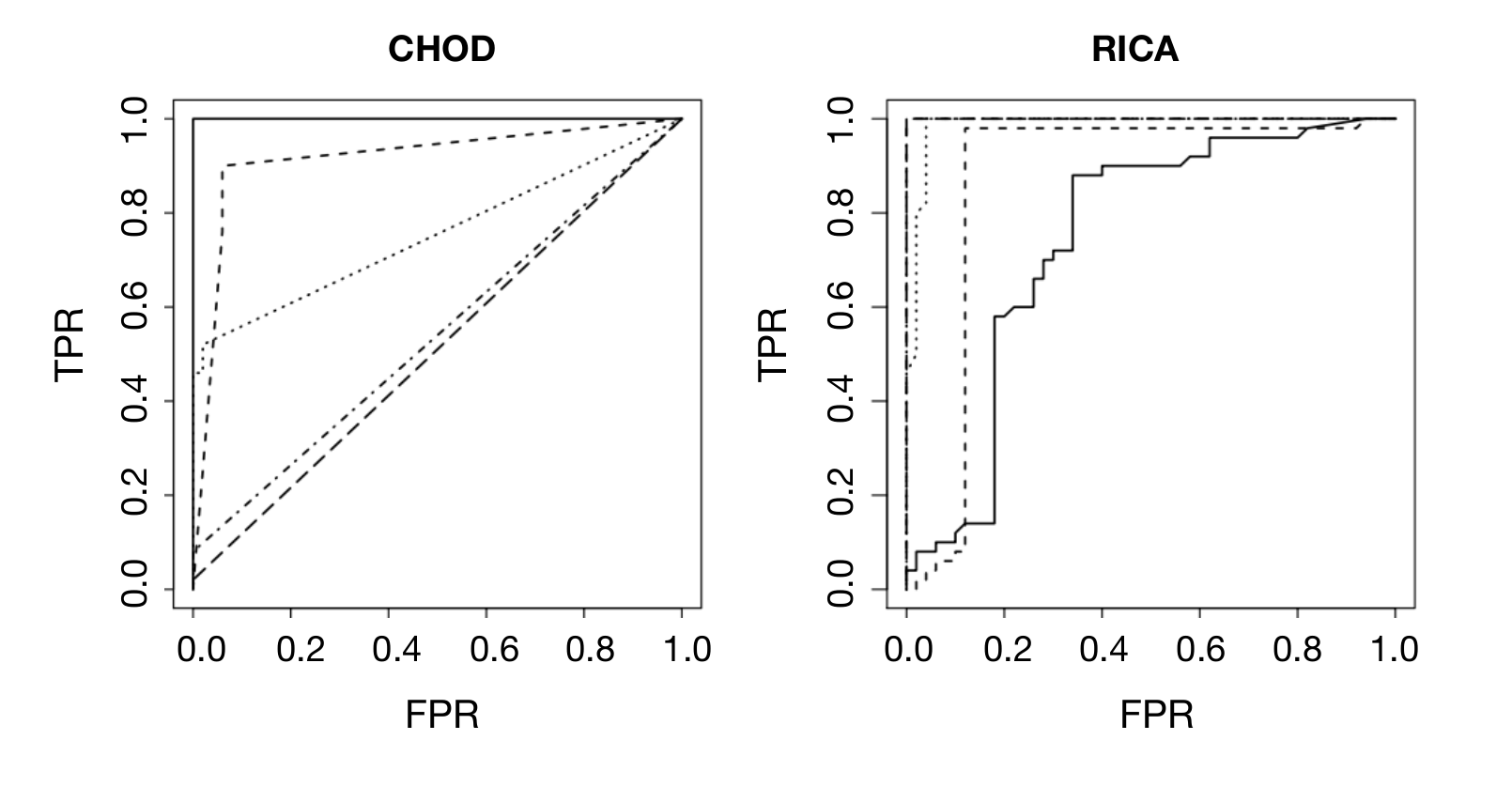}
\caption{Three-node graph with uniform noises.}
\end{subfigure}
\begin{subfigure}[b]{0.46 \textwidth}
\centering
\includegraphics[width = 1 \textwidth]{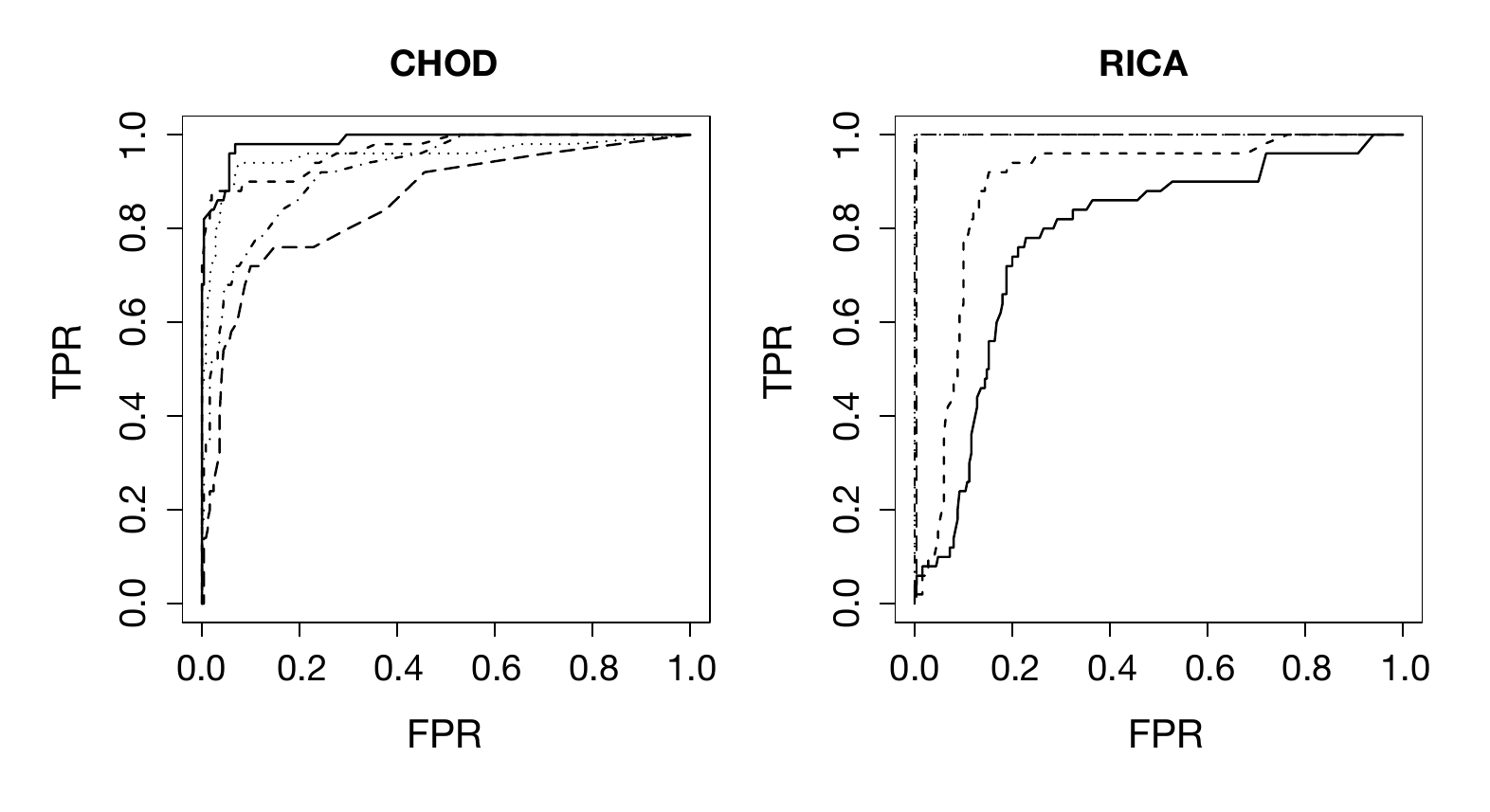}
\caption{Four-node graph with uniform noises.}
\end{subfigure}
\caption{Misspecification 1. (a) Simulation true direct causal effect functions. (b)\&(c) Simulation true graphs. Solid red nodes are latent (discarded at model fitting stage). (d)\&(e) Receiver operating characteristics curves for recovering causal relationships between observed variables under varying degrees of heterogeneity are represented by the same line types as shown in (a).}
\label{misspecified}
\end{figure*}

\paragraph{Misspecification 1: non-linear confounding, non-Gaussianity, and varying degrees of heterogeneity} From previous experiments, the proposed CHOD consistently outperformed non-Gaussian SEMs because data were heterogeneous and the errors were Gaussian, both conditions favoring CHOD. For fairer comparison and better illustration, we conducted further simulations under an alternative data generating mechanism. Specifically, we mimicked the simulation setting in \citet{salehkaleybar2020learning} by generating $n = 250$ observations from the SEM \eqref{sem} with uniform noises $\varepsilon\sim U(-1, 1)$ under a three-node (Figure \ref{misspecified}(b)) and a four-node (Figure \ref{misspecified}(c)) directed acyclic graph, both having one unmeasured confounder (red, discarded at the model fitting stage). The non-zero direct causal effects were assumed to be quadratic with varying degrees of curvature (Figure \ref{misspecified}(a)). As the degree of curvature approached zero, the data became more homogeneous. The coefficient functions were scaled to have the same average effects, that is we kept $\int |b(Z)| dZ$ constant.

We compared CHOD with RICA. Their receiver operating characteristic (ROC) curves under each true graph are shown in Figure \ref{misspecified}(d) and (e), respectively. In both cases, CHOD's performance deteriorated as the degree of heterogeneity \emph{decreased}; by contrast, RICA's performance deteriorated as the degree of heterogeneity \emph{increased}. Not surprisingly, when the data were completely homogeneous, CHOD was no better than a random guess in the three-node graph. However, somewhat surprisingly, in the four-node graph, CHOD still performed reasonably well even when the data were homogeneous. For instance, the worst area under the ROC curve were 0.865 and 0.789 for CHOD and RICA, respectively. Also note that, given sufficient degrees of heterogeneity, CHOD performed reasonably well despite the non-Gaussianity and non-linear confounding.

For further comparison, we reran the analyses on data generated with Gaussian noises (keeping everything else the same). The results are shown in Section S3.2 of the Supplementary Materials. RICA were close to random guesses in both graphs whereas CHOD thrived on heterogeneity, especially in the four-node graph.

\paragraph{Misspecification 2: partially homogeneous data and unknown covariates} We considered another type of model misspecification where data were partially homogeneous (i.e., observations were clustered and iid within each cluster) and the unknown covariate was estimated by UMAP. CHOD significantly outperformed the competing methods (see Section S3.2 of the Supplementary Materials).

\subsection{An Application to Breast Cancer Genomic Data}

We demonstrate the capability of CHOD in identifying gene feedback loops using breast cancer gene expression data downloaded from the Cancer Genome Atlas (\url{https://www.cancer.gov/tcga}). Breast cancer is a well-known extremely heterogeneous genetic disease. The dataset contains $n = 1215$ observations with 113 normal and 1102 tumor tissues. We focused on 8 feedback loops involving gene TP53 \citep[][plotted in Section S3.3 of the Supplementary Materials]{harris2005p53}. We compared CHOD with two cyclic causal discovery methods, LiNG and ANM. In addition, we also considered two versions of JCI \cite{mooij2020joint}, FCI-JCI and ASD-JCI, which are broadly applicable for causal discovery with heterogeneous data. Gene expressions were log-transformed. For CHOD and JCI, we learned a one-dimensional embedding using UMAP as an input covariate and regressed out the effects of the covariate on the mean gene expression. The results are reported in Table \ref{application}. CHOD had uniformly strong performance: it was only outperformed by LiNG in one case in terms of MCC. See Section S3.3 of the Supplementary Materials for more elaborated description of the comparison.

\begin{table}[h]
\caption{Application to breast cancer genomic data. 8 feedback loops involving gene TP53 were considered. The best performance is shown in boldface. } \label{application}
\begin{center}
\resizebox{1\textwidth}{!}{\begin{tabular}{cccccccccc}
\toprule
\multirow{2}{*}{Method} & \multicolumn{3}{c}{Network A} & \multicolumn{3}{c}{Network B} & \multicolumn{3}{c}{Network C} \cr
\cmidrule(lr){2-4} \cmidrule(lr){5-7} \cmidrule(lr){8-10}
& TPR & FDR & MCC & TPR & FDR & MCC & TPR & FDR & MCC \cr
\midrule
CHOD & \textbf{0.750} & 0.250 & 0.550 & 0.333 & \textbf{0.200} & \textbf{0.398} & \textbf{0.667} & \textbf{0.500} & \textbf{0.316} \cr
LiNG & \textbf{0.750} & \textbf{0.000} & \textbf{0.791} & 0.583 & 0.563 & 0.198 & 0.333 & 0.667 & 0.000 \cr
ANM & 0.500 & 0.333 & 0.316 & \textbf{0.667} & 0.556 & 0.238 & 0.333 & \textbf{0.500} & 0.189 \cr
FCI-JCI & 0.250 & 0.500 & 0.059 & 0.417 & 0.500 & 0.219 & \textbf{0.667} & \textbf{0.500} & \textbf{0.316} \cr
ASD-JCI & 0.500 & 0.333 & 0.316 & 0.500 & 0.455 & 0.298 & \textbf{0.667} & \textbf{0.500} & \textbf{0.316} \cr
\midrule
\multirow{2}{*}{Method} & \multicolumn{3}{c}{Network D} & \multicolumn{3}{c}{Network E} & \multicolumn{3}{c}{Network F} \cr
\cmidrule(lr){2-4} \cmidrule(lr){5-7} \cmidrule(lr){8-10}
& TPR & FDR & MCC & TPR & FDR & MCC & TPR & FDR & MCC \cr
\midrule
CHOD & \textbf{0.500} & 0.571 & \textbf{0.275} & \textbf{0.833} & \textbf{0.286} & \textbf{0.693} & 0.600 & \textbf{0.250} & \textbf{0.545} \cr
LiNG & \textbf{0.500} & 0.667 & 0.164 & 0.333 & 0.778 & 0.031 & 0.800 & 0.500 & 0.405 \cr
ANM & 0.167 & \textbf{0.500} & 0.180 & 0.667 & 0.556 & 0.359 & \textbf{1.000} & 0.583 & 0.389 \cr
FCI-JCI & 0.333 & 0.667 & 0.123 & 0.500 & 0.700 & 0.114 & 0.400 & 0.667 & 0.035 \cr
ASD-JCI & 0.167 & 0.750 & 0.010 & 0.500 & 0.625 & 0.217 & 0.600 & 0.571 & 0.221 \cr
\midrule
\multirow{2}{*}{Method} & \multicolumn{3}{c}{Network G} & \multicolumn{3}{c}{Network H} \cr
\cmidrule(lr){2-4} \cmidrule(lr){5-7}
& TPR & FDR & MCC & TPR & FDR & MCC \cr
\midrule
CHOD & \textbf{1.000} & \textbf{0.000} & \textbf{1.000} & \textbf{1.000} & \textbf{0.000} & \textbf{1.000} \cr
LiNG & \textbf{1.000} & \textbf{0.000} & \textbf{1.000} & 0.500 & 0.000 & 0.577 \cr
ANM & \textbf{1.000} & \textbf{0.000} & \textbf{1.000} & 0.500 & 0.000 & 0.577 \cr
FCI-JCI & 0.000 & - & - & 0.000 & - & - \cr
ASD-JCI & 0.000 & - & - & 0.000 & - & - \cr
\bottomrule
\end{tabular}}
\end{center}
\end{table}

\section{DISCUSSION}
We have proposed one of the first model-based identifiable causal discovery methods for observational data in the presence of both cyclic causality and confounders by exploiting the heterogeneity of causal mechanism, under which most existing methods fail. Our model formulation can be justified from two perspectives. First, in the regression context, the varying-coefficient model serves as an important generalization of (generalized) linear model. As introduced by \cite{hastie1993varying}, the class of varying-coefficient models ties together many important structured regression models such as additive models and dynamic linear models into one common framework. Likewise, our proposed model is a natural extension of linear SEMs, which allows varying causal effects. One important ingredient exploited in this paper is that the adoption of varying causal effects helps identify the causal structure. Second, if we restrict the covariate to be categorical, our model reduces to multi-domain/group-specific graphical models (see, for example, \cite{yajima2015detecting, ghassami2018multi, ni2018reciprocal}), which have been proved successful in many applications.

There are many additional future directions can be taken to extend this work. For example, we have focused on linear Gaussian models which can be extended for nonlinear and non-Gaussian data, although our simpler model does have some advantages that nonlinear/non-Gaussian models do not share; see the brief discussion of the inference of causal effects in Section S4 of the Supplementary Materials. In our current implementation, we assume the causal effects can vary with covariates but the causal structure does not. Our theories are directly applicable but for implementation we need to impose spike-and-slab prior on each individual spline coefficient rather than the spline coefficient vector. 

\clearpage

\bibliography{reference}

\end{document}